# STS study of the CMR effect of a manganite thin film in an external magnetic field

B. Damaschke, T. Mildner, V. Moshnyaga and K. Samwer

I. Physikalisches Institut, Georg-August-Universität Göttingen, Friedrich-Hund-Platz 1, Göttingen, Germany

E-mail: bdamasc1@gwdg.de

A $La_{0.75}Ca_{0.25}MnO_3$-film grown by metalorganic aerosol deposition technique was investigated by scanning tunnelling microscopy and spectroscopy. A small spot was found on the surface which exhibits the expected magnetic field dependence of the tunnelling conductivity giving the opportunity for a local spectroscopic study of the intrinsic colossal magnetoresistance (CMR) behavior. The tunnelling conductivity is strongly enhanced in an external magnetic field of 4 T and the CMR behavior can be interpreted in terms of a redistribution of occupied electronic states torwards the Fermi energy.

**Introduction**

The colossal magnetoresistance (CMR) effect in doped manganites is in most cases accompanied by a metal to insulator (MI) transition [1 - 3]. The physics of the transition is marked by the interplay of spin, charge, lattice, and orbital degrees of freedom leading to a rich phase diagram with electrically insulating and metallic phases with different magnetic behavior [4]. Despite the experimental and theoretical efforts made in the last years no generally accepted theory or microscopic model has been established until now [5 - 8].

With the help of scanning tunneling microscopy/spectroscopy (STM/STS) it was possible to distinguish between the different phases locally due to their electrical properties [9 - 12]. Especially STS measurements in external magnetic fields can give information about the phase distribution and the dependence on the topography of the surface and the local CMR behavior [9, 13]. Phase separation as well as continuous transitions were found in manganite films depending on the details of the samples and the doping level [13, 14].

In this work a thin $La_{0.75}Ca_{0.25}MnO_3$-film was characterized with X-ray diffraction, magnetization measurements (SQUID) and electric transport in an external magnetic field. The sample was investigated in detail with STM/STS in UHV and in an external magnetic field in the vicinity of the MI-transition temperature. A similar film showing cation ordering and a very sharp transition was used in Ref. 15 where we investigated the structure and the development of the physical properties as a function of temperature. In this work we focused on STM/STS measurements under the influence of external magnetic fields.

**Experimental results**

The sample is a thin $La_{0.75}Ca_{0.25}MnO_3$-film grown by Metalorganic Aerosol Deposition (MAD) technique [16] on a MgO substrate. It is fully epitaxial with a lattice constant of 0.3872 nm (rhomboedric structure). The thickness of 80 nm was evaluated from wide angle X-ray diffraction experiments. In Fig. 1 the resistivity, CMR and magnetization is shown as a function of temperature. The MI-transition temperature is 252 K, the Curie temperature is 258 K as measured in a SQUID magnetometer (criterion: maximum of slope). The CMR has its maximum around the MI-transition. These results should be compared to those of a similar film in Ref. 15 where a fully cation ordered film was investigated showing an extremely sharp transition. A comparison to these data leads to the interpretation that the film investigated here is partially cation ordered (smaller transition temperature, transition width larger).

The STM/STS measurements were performed in a Cryo-UHV-scanning probe system (Omicron NanoTechnology GmbH) equipped with a superconducting magnet coil (max. 7 T, field perpendicular to surface).

The STM investigations on the sample are difficult because of the roughness of the surface and the fact that the sample was prepared not in situ but with MAD technique under air. This may lead to surface reconstruction and partial surface contamination. These difficulties cannot be circumvented by in situ techniques due to problems with the oxygen stoichiometry and therefore the surface has to be investigated intensively to find suitable parts that show the undisturbed behavior of the CMR manganites. In Fig. 2 such a surface area is shown. In the topography (top of Fig. 2) a spot was found (inside the white box) which shows a very smooth surface structure while the surrounding parts show a typical ripple structure which is found on most parts of the sample surface. The measurements were made without and in an external perpendicular magnetic field of 4 T at a temperature of 238 K (below the MI-transition). In the topography no changes were found in the magnetic field, as expected. The sample was also investigated by STS. U-I-characteristics were measured every 2 nm in each line. From the U-I-characteristics the slope dI/dU at 0 V was evaluated by a linear fit around 0 V in the region where the curve is linear. The slope was used to plot the STS results in the center of Fig. 2 (tunnelling conductance at 0 V was used for the color scale). The spot shows the expected behavior of a CMR manganite bulk sample: In zero field it is smooth and the local tunnelling conductance is the same as in the surrounding matrix. In a field of 4 T, however, the tunnelling conductance changed drastically in the spot whereas the changes in the matrix are only marginal. The histogram of the tunnelling conductance values in the white box is shown in the bottom of Fig 2. Without field (red curve) a broad distribution occurs while in a field of 4 T the distribution (blue curve) splits into two maxima for the spot (right maximum at 0.108 nA/V) and the matrix (left maximum at 0.077 nA/V). The difference is about 40 % which is comparable to the transport measurements of the resistivity at 238 K. So there is a clear field effect of the spot together with a qualitative agreement with the CMR behavior in the transport measurement. From this behavior it can be concluded that we have a clean spot where we can investigate the intrinsic behavior of the manganite sample.

The details of the characteristics are shown in Fig. 3. We have chosen two distinct parts of Fig. 2, one is the spot ('spot') and the other is a typical part of the matrix of the same width ('matrix'). The derivative of the characteristics of these areas (mean values) are shown in Fig 3. For the spot we found an increase of the tunnelling conductance due to the magnetic field mainly for negative bias voltages that means tunnelling from the sample to the tip (filled states) and also in the vicinity of the Fermi energy up to 0.1 V. For positive voltages above 0.1 V no effect was seen. In the matrix no significant changes were detected, for negative voltages the dI/dU is much steeper compared to the spot. That means that the tunnelling current is much stronger for the matrix than for the spot in this voltage range. This leads to the local voltage dependence of the tunnelling conductance shown in Fig 4. Without an external magnetic field the spot is dark for large negative voltages and the contribution of the matrix to the tunnelling current dominates. For increasing voltage this effect is compensated. At 4 T (right side of Fig. 4) the spot appears only near the Fermi energy.

**Discussion**

From the topography data it can be seen that the spot is a small region at the flank of a bigger terrace. The surface is much smoother and the noise smaller than on the surrounding matrix. The data suggest that the spot has a cleaner surface compared with the matrix. Another explanation may be that the spot exhibits an unstrained microstructure and therefore the electronic properties behave undisturbed.

The dI/dV curves of the spot and the matrix in Fig. 3 are asymmetric and shifted about 30 mV to the left with respect to the Fermi energy (0 V). This effect can be attributed to a difference in the work functions of tip and sample ($\Phi_{sample} < \Phi_{tip}$) [17]. A detailed analysis (fit around the minimum) shows that we get the following values for the position of the minimum, dI/dV$_{min}$: spot(0T) = -42 mV, spot(4T) = -35 mV, matrix = -30 mV. That means that the work function of the spot increases in a magnetic field (shift of the minimum by 7 mV). This behavior is in qualitative agreement with measurements of the tunnelling current as a function of distance (I(d)-spectroscopy, not shown) near the MI-transition temperature.

The slope of the dI/dV-curves for negative voltages is rather different for the spot and the matrix: The dI/dV of the matrix is much steeper whereas for positive voltages the behavior is more or less the same for both regions. This is illustrated in Fig. 4 by the development of the spot as a function of voltage in comparison with the matrix. The spot exhibits the typical CMR-behavior only in the vicinity of the Fermi energy. In the picture of the density of states this means that in the matrix the density of the occupied states is much more pronounced than in the spot. This lead to the suggestion that a transition from an insulating (matrix like) 'phase' to a metallic (spot like) 'phase' is accompanied by a redistribution of the density of occupied states towards the Fermi energy. The enhancement of dI/dV of the spot in the external magnetic field is only substantial for negative voltages and therefore the CMR effect is leading to a change only of the occupied states.

Both regions show a bending of their dI/dV-curves at +25 – +80 mV which becomes more pronounced for the spot in a magnetic field. For higher voltages the curves are not distinguishable so that the density of states is the same for both 'phases' in this region (unoccupied states).

A comparison of the magnetic field driven transition with the temperature driven transition (STS data not shown here but are similar to those discussed in detail in Ref. 11) shows a qualitative agreement: in both cases the STS data show a drop in the zero-bias tunnelling conductance and a development towards a gap-like behavior when the metallic phase is weakened by higher temperatures or lower magnetic fields.

**Conclusion**

In summary, the CMR behavior of a LCMO film was investigated by STM/STS. We were able to find a spot on the surface showing the intrinsic behavior for doped manganites. The STS results indicate that the CMR effect is accompanied by a redistribution of the density of occupied states towards the Fermi energy. Occupied states are not involved.

**Acknowledgements**

We greatfully acknowledge funding by the DFG via SFB 602.

**Figure captions**

**Fig. 1:** Resistivity ρ of the LCMO film as a function of temperature with and without magnetic field. Inset left: CMR as a function of temperature. Inset right: Magnetization as a function of temperature.

**Fig. 2:** STM/STS results on LCMO film. Top: topography (left) with sectional view (right) of the spot along the blue line. Center: STS without magnetic field (left) and in 4 T (right). Bottom: Histogram of tunnelling conductance σ (only data in white box) without (red) and in 4 T (blue): Splitting of the distribution into two maxima in magnetic field.

**Fig. 3:** Tunneling conductance versus voltage (data from the white box in Fig. 1) of the spot (filled symbols) and the matrix (open symbols) with (blue) and without (red) magnetic field.

**Fig. 4:** STS with (right) and without (left) magnetic field as a function of bias voltage.

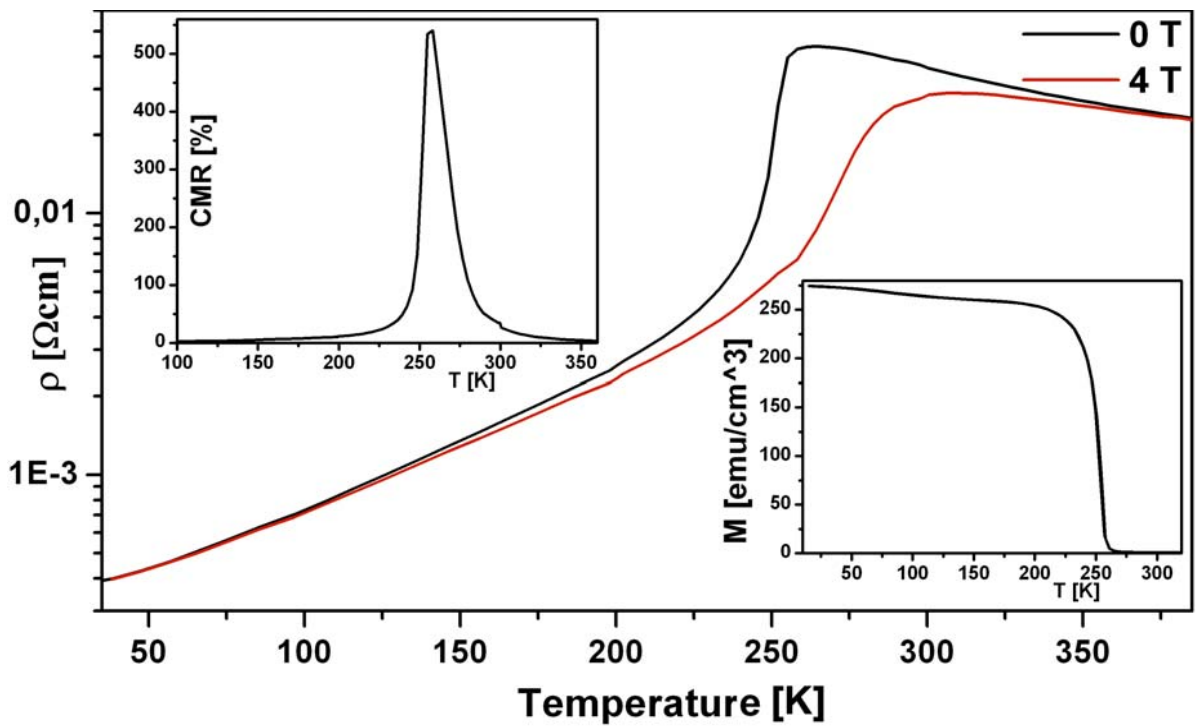

**Fig. 1:** Resistivity ρ of the LCMO film as a function of temperature with and without magnetic field. Inset left: CMR as a function of temperature. Inset right: Magnetization as a function of temperature.

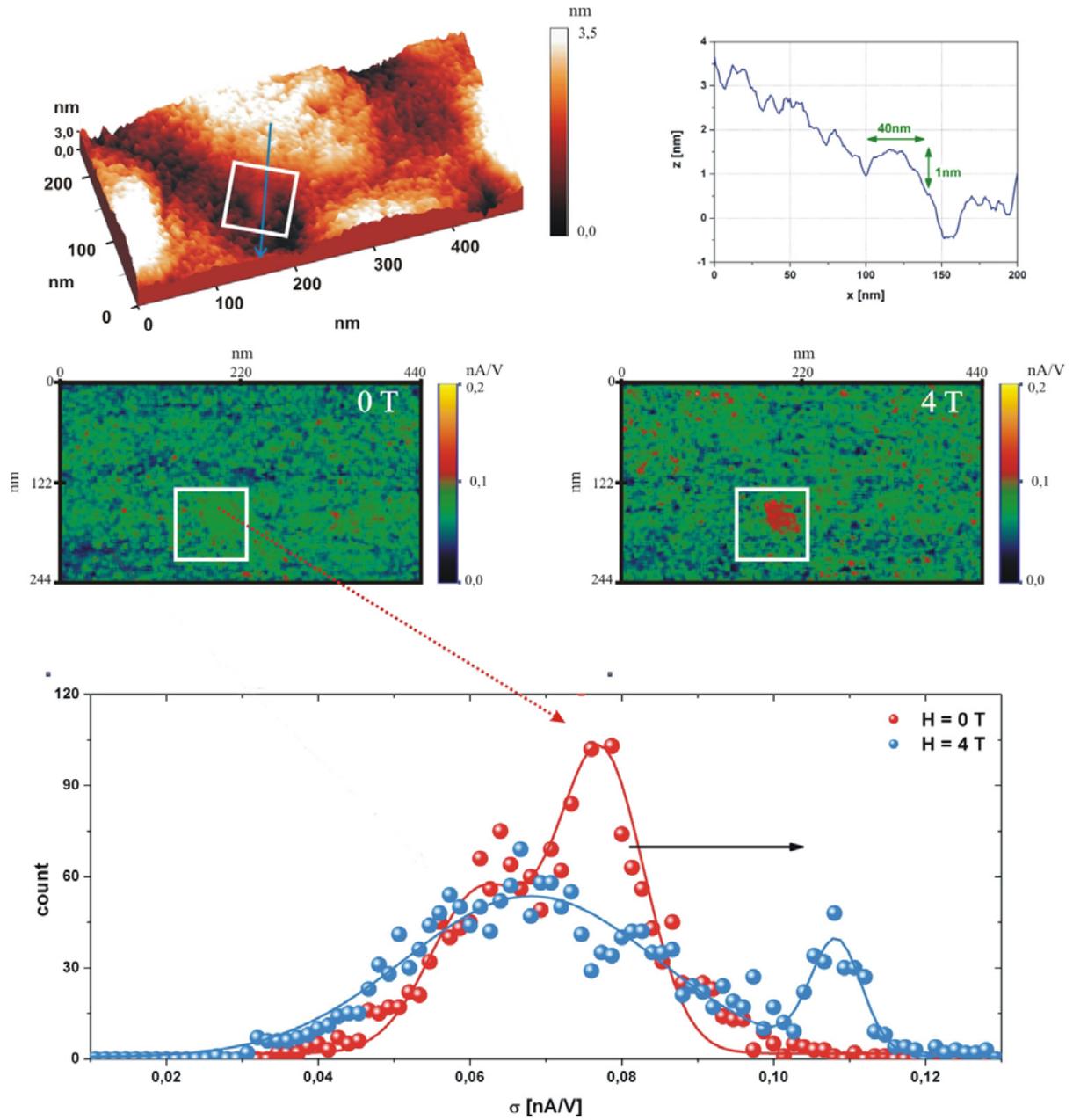

**Fig. 2:** STM/STS results on LCMO film. Top: topography (left) with sectional view (right) of the spot along the blue line. Center: STS without magnetic field (left) and in 4 T (right). Bottom: Histogram of tunnelling conductance σ (only data in white box) without (red) and in 4 T (blue): Splitting of the distribution into two maxima in magnetic field.

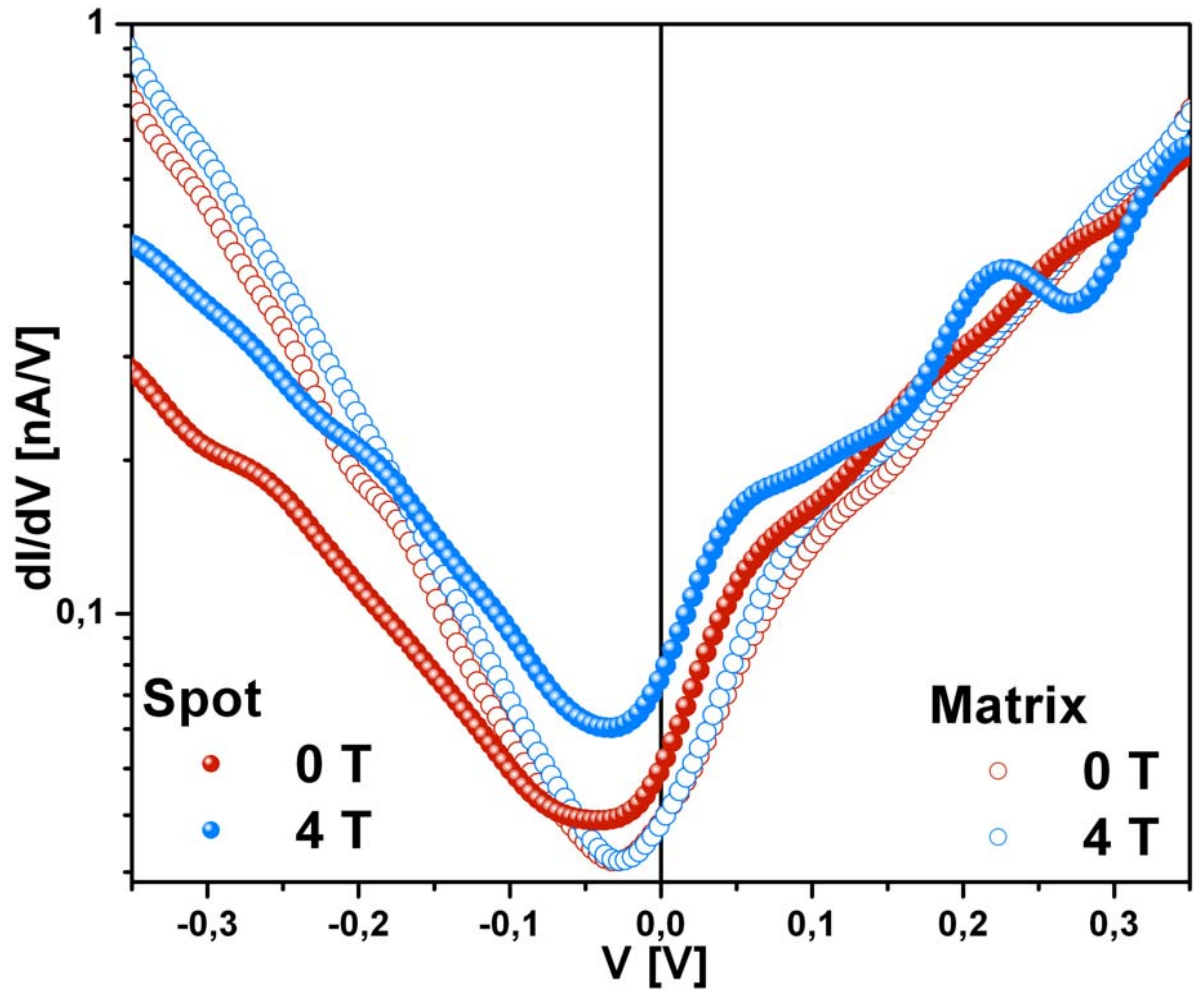

**Fig. 3:** Tunneling conductance versus voltage (data from the white box in Fig. 1) of the spot (filled symbols) and the matrix (open symbols) with (blue) and without (red) magnetic field.

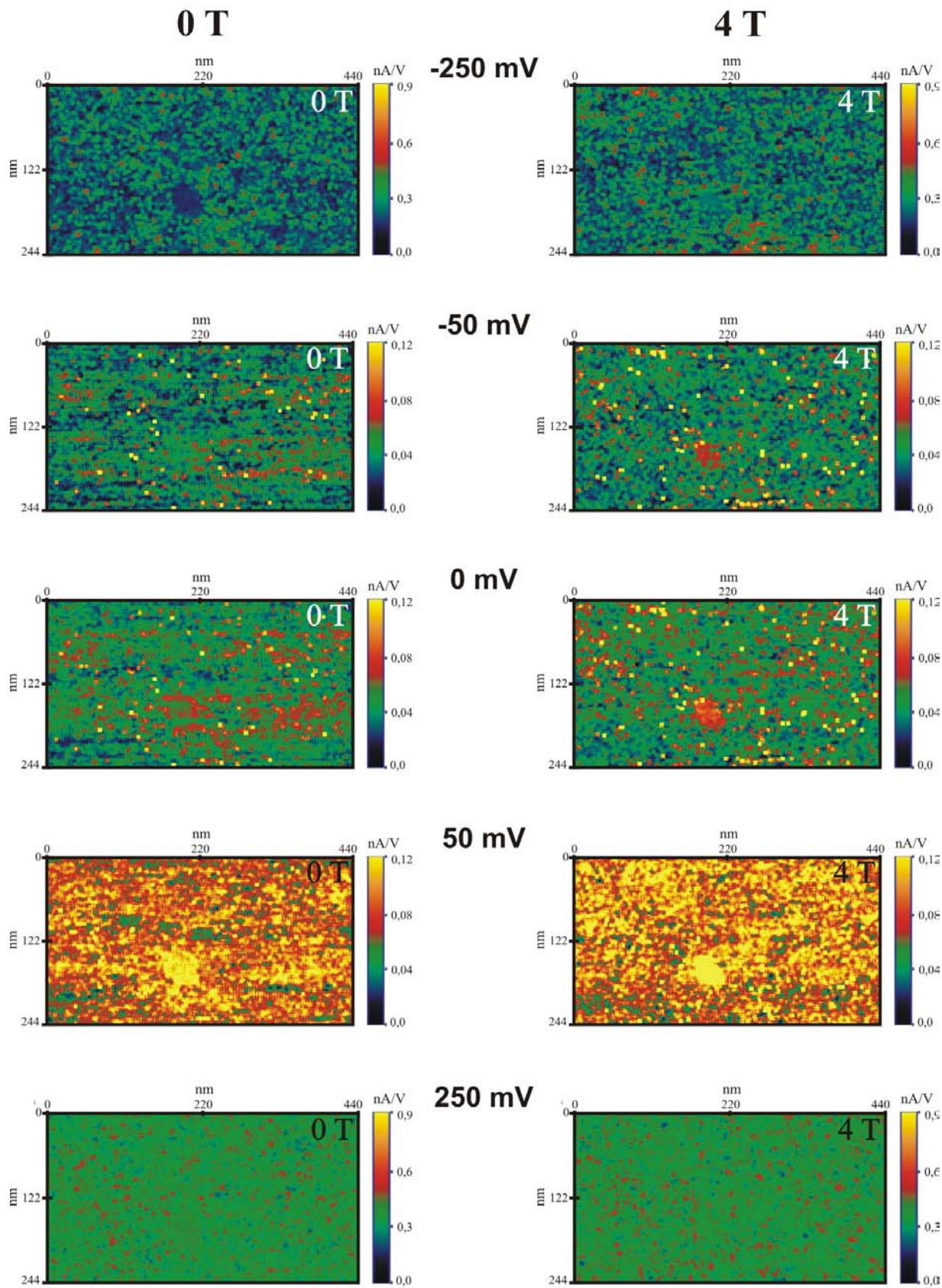

**Fig. 4:** STS with (right) and without (left) magnetic field as a function of bias voltage.